\documentclass[12pt]{article}
\usepackage{graphics,color}
\begin{document}
\thispagestyle{empty}
%\pagestyle{empty}
%\vspace*{2.5cm}
\noindent\
\\
\\
\\
\begin{center}
\large \bf  Composite Weak Bosons at the LHC
\end{center}
\hfill
 \vspace*{1cm}
\noindent
\begin{center}
{\bf Harald Fritzsch}\\
Department f\"ur Physik\\ 
Ludwig-Maximilians-Universit\"at\\
M\"unchen, Germany \\

\vspace*{0.5cm}
\end{center}

\begin{abstract}

In a composite model of the weak bosons  
the p-wave bosons are studied.
The state with the lowest mass is identified with the 
boson, which has been observed at the LHC. Specific properties of 
the excited bosons 
are studied, in particular their decays into weak bosons and photons.
Such decays might have been observed recently with the ATLAS detector 
at the Large Hadron Collider.  

\end{abstract}

\newpage

Three years ago a scalar boson with a mass of about 125 GeV has been discovered at CERN (ref. (1)).  
This boson might be an excited weak boson (see ref. (2)).\\

Recently one has observed effects, which indicate the 
existence of particles with a mass of about 2 TeV, decaying into two weak bosons (ref. (3)). These 
particles could be the excited weak bosons, discussed in ref. (2). They decay into two weak bosons, which are observed as two high energy jets.\\ 

In this letter we discuss the specific properties of excited weak bosons. They consist as the weak bosons of a fermion and its antiparticle, which are denoted as "haplons". They are bound by the gauge bosons of the new confining gauge theory "Quantum Haplo Dynamics" ($QHD$). The $QHD$ mass scale is given by a mass parameter $\Lambda_h$, which determines the size of the weak bosons.\\

Two types of haplons are needed as constituents of the weak bosons, denoted by $\alpha$ and $\beta$.  
Their electric charges in units of e are:

\begin{equation}
Q = \left( \begin{array}{l}
+\frac{1}{2}\\
-\frac{1}{2}\\
\end{array} \right) \ .
\end{equation}
\\
The three weak bosons have the following internal structure:
\begin{eqnarray}
W^+ & = & \overline{\beta} \alpha \; , \nonumber \\
W^- & = & \overline{\alpha} \beta \; , \nonumber \\
W^3 & = & \frac{1}{\sqrt{2}} \left( \overline{\alpha} \alpha -
\overline{\beta} \beta \right) \; .
\end{eqnarray}

In the absence of electromagnetism the weak bosons are degenerate in mass. If the electromagnetic interaction is introduced, the mass of the neutral boson increases due to the mixing with the photon (see ref. (2)).\\

The $QHD$ mass scale is about thousand times larger than the $QCD$ mass scale. In strong interaction physics 
above the energy of 1 GeV many resonances exist. We expect similar effects in the electroweak sector. At high energies there should exist excited weak bosons, which decay mainly into two or three weak bosons. Such states might have been observed at the $LHC$ (ref. (3)).\\

The weak bosons consist of pairs of haplons, which are in an s-wave. The spins of the two 
haplons are aligned, as the spins of the quarks in a $\rho$-meson. The first excited states are those, in 
which the two haplons are in a p-wave. We describe the quantum numbers of these states by $I(J)$. 
The $SU(2)$-representation is denoted by $I$, $J$ describes the total angular momentum.\\ 

There are three $SU(2)$-singlets, which we denote by S:
\begin{eqnarray}
 S=& \frac{1}{\sqrt{2}} \left( \overline{\alpha} \alpha +
\overline{\beta} \beta \right) \ .
 \end{eqnarray}
\begin{eqnarray}
S(0)=[0(0)], \nonumber \\
S(1)=[0(1)], \nonumber \\
S(2)=[0(2)]. 
\end{eqnarray}

The three $SU(2)$ triplets are denoted by T:
\begin{eqnarray}
T^+ & = & \overline{\beta} \alpha \; , \nonumber \\
T^- & = & \overline{\alpha} \beta \; , \nonumber \\
T^3 & = & \frac{1}{\sqrt{2}} \left( \overline{\alpha} \alpha -
\overline{\beta} \beta \right) \; .
\end{eqnarray}
\begin{eqnarray}
 T(0)=[1(0)],  \nonumber \\
 T(1)=[1(1)], \nonumber \\
 T(2)=[1(2)].
\end{eqnarray}
 
The boson $S(0)$ is identified with the particle, observed at CERN (ref. (1)). Thus the mass of $S(0)$ is about 125 GeV. The masses of the eleven p-wave states  $S(1),S(2)$ and  $T(0),T(1),T(2)$ are expected to be above 0.3 TeV. In analogy to $QCD$ we expect, that the masses of the $T$ - bosons are larger than the masses of the corresponding $S$-bosons. The mass of  the $T(2)$-boson might be as large as 2 TeV.\\

The excited weak bosons will be produced mainly by the fusion of two weak bosons, emitted from the quarks. The production rates cannot be calculated, since they depend on details of the wave functions.\\ 

The $S(1)$ - boson can  decay into two weak bosons, but also into three or four weak bosons. Let us consider first the decay into two weak bosons. The $Z$-boson is a mixture of the boson $W^3$ and the photon. The mixing angle is 
the weak angle, measured to about 28.7 degrees.\\

Using this angle, we can calculate the 
branching ratios BR for the decays into two weak bosons, into a weak boson and a photon and into two photons. The branching ratio BR for the decay into two charged weak bosons is denoted by 2B. We obtain:\\
\\
$S(1) \Longrightarrow  W^+ + W^- $\\              
(BR $\simeq 2B $),\\
$S(1) \Longrightarrow  Z + Z $\\                      
(BR $\simeq 0.59 ~B $), \\
$S(1) \Longrightarrow   Z +  \gamma  $\\           
(BR $\simeq 0.35 ~B $), \\
$S(1) \Longrightarrow  \gamma + \gamma $\\    
(Br$\simeq 0.05 ~B$). \\ 

The boson $S(2)$ will decay mostly into three or four weak bosons or photons, e.g.: \\
\\
$S(2) \Longrightarrow  W^+ + W^- + Z $,\\
$S(2) \Longrightarrow  W^+ + W^- + W^+ + W^- $,\\
$S(2) \Longrightarrow  W^+ + W^- + \gamma $.\\
\\
It can also decay into the S(1)-boson and a Z-boson or a photon:\\
\\
$S(2) \Longrightarrow  S(1) + Z $,\\
$S(2) \Longrightarrow  S(1) + \gamma $.\\

The $SU(2)$ - triplet bosons $T(0)$, $T(1)$ and $T(2)$ will decay mainly into two or three weak bosons or photons. For example, interesting decay modes of $T(0)^+$ would be:\\
\\
$T(0)^+ \Longrightarrow  (W^+ + Z + Z) $,\\
$T(0)^+ \Longrightarrow  (W^+ + Z + \gamma) $,\\
$T^(0)+ \Longrightarrow  (W^+ + \gamma + \gamma) $.\\

It is very difficult to observe the decays into more than two weak bosons or photons, but the decays into two weak bosons or photons can be observed, for example the decays:\\ 
\\
$T(0)^+ \Longrightarrow  (W^+ + Z) $,\\
$T(0)^+ \Longrightarrow  (W^+ + \gamma) $.\\ 

At the LHC one has observed recently effects, which might be due to the decays of the 
S(2) or T(2) bosons with a mass of about 2 TeV, decaying into two weak bosons (ref.(3)). The decay 
into two weak bosons can be observed by measuring the energies and momenta of two high energy jets, 
which are due to the decays of the W- or Z-bosons: 
\\
\\
$T(2)^+ \Longrightarrow  (W^+ + Z) $,\\
$T(2)^0 \Longrightarrow  (W^+ + W^-) $,\\
$T(2)^0 \Longrightarrow  (Z + Z) $,\\
$T(2)^- \Longrightarrow  (W^- + Z) $,\\ 
$S(2)   \Longrightarrow  (Z + Z) $,\\
$S(2)   \Longrightarrow  (W^+ + W^-) $.\\

The three bosons $T(2)^+$, $T(2)^0$ and $T(2)^-$ are degenerate in the absence of electromagnetism - the electromagnetic splitting of the masses is of the order of 1 GeV. It is expected that the mass of the S(2)-boson is less than the mass of the T(2)-boson. If we use the analogy with $QCD$, the mass splitting is estimated to about 60 GeV.\\

We consider the decays of the S(2)-boson into two weak bosons, into one weak boson and a photon and into two photons:\\
\\
$S(2)\Longrightarrow (W^+ + W^-)$\\
(BR: 2B),\\ 
$S(2)\Longrightarrow  (Z + Z) $\\ 
(BR:  $cos^4\theta \times B \simeq 0.59 B $, $\theta$: weak mixing angle),\\
$S(2) \Longrightarrow  (Z + \gamma) $\\ 
(BR: $ 2 sin^2\theta \times cos^2\theta \times B \simeq 0.35 B $),\\ 
$S(2) \Longrightarrow  (\gamma + \gamma) $\\ 
(BR: $sin^4\theta \times B \simeq 0.05 B $).\\

Thus five decays of S(2) into two Z-bosons are accompanied by three decays of S(2) into a Z-boson and a photon - twelve decays of S(2) into two Z-bosons are accompanied by one decay of S(2) into two photons.\\

Now we consider the decays of the charged T(2)-bosons  - the decays of the neutral T(2)-boson are analogous to the decays of the S(2)-boson): \\
\\
$T(2)^+ \Longrightarrow  (W^+ + Z) $\\ 
(BR: $cos^2\theta \times B \simeq 0.77B $),\\
$T(2)^+ \Longrightarrow  (W^+ + \gamma) $\\ 
(BR: $sin^2\theta \times B \simeq 0.23B) $.\\
\\
Thus ten decays of  $T(2)^+$ into a W-boson and a Z-boson are accompanied by three decays into a W-boson and a photon. The branching ratio B cannot be calculated -  it depends on how often the S- or T-bosons decay into three or more bosons - we estimate $B \simeq 0.30$.\\

The decays of S(2) and T(2) into two weak bosons, which decay into quarks, produce two high energy jets. These jets have  been perhaps observed recently (see ref. (3)). But in this case also the decays into one weak boson and a photon and the decays into two photons must soon be observed.\\

Interesting are the decays of S(2) and T(2) into two Z-bosons, which subsequently decay into muons: 
\\

$S(2)\Longrightarrow  Z + Z \Longrightarrow (\mu^+\mu^-)+(\mu^+\mu^-). $\\
\\
One should observe four muons as decay products of two Z-bosons. The ratio of the decay rates of S(2) into four muons and into hadrons can be calculated:
\\
\begin{equation}
\frac{rate(S(2)\Longrightarrow  Z + Z \Longrightarrow (\mu^+\mu^-)+(\mu^+\mu^-))}{rate(S(2)\Longrightarrow  Z + Z \Longrightarrow (hadrons)+(hadrons))} \simeq 0.0023.\\
\end{equation}
\\
This ratio is very small. For each decay of S(2) into four  Z-bosons muons there are about 430 decays of (Z,Z) into hadrons. But the signal is very clear - if enough data are available, the decays into four muons should be observed.\\

Furthermore one should try find the events, where one Z-boson decays into quarks and the second Z-boson decays into muons:
\\
\begin{equation}
\frac{rate(S(2)\Longrightarrow  Z + Z \Longrightarrow (\mu^+\mu^-)+(hadrons))}{rate(S(2)\Longrightarrow  Z + Z \Longrightarrow (hadrons)+(hadrons))} \simeq 0.05.\\
\end{equation}
\\
Also the decays of S(1) and of T(0) and T(1) into two weak bosons should be observed. But this is more difficult, since the energies of the two weak bosons are less than the energies of the weak bosons in the S(2) or T(2) decays. If the masses of the S(2) and T(2)-boson are about 2 TeV, the masses of the S(1)- and T(1)-bosons should be about one TeV.\\     

Thus the Large Hadron Collider has good chances to discover another layer of substructure in our universe.

\end{document}